\documentstyle[12pt, aaspp4]{article}
 
\def\kms{km s${}^{-1}$}

\def\a{$\alpha$}

\def\m{$\mu$m}
\def\ab{$\sim$}
\def\p{$\pm$}
\def\cii{[C\kern.2em{\sc ii}]}
\def\oi{[O\kern.2em{\sc i}]}
\def\oiii{[O\kern.2em{\sc iii}]}
\def\nii{[N\kern.2em{\sc ii}]}
\def\niii{[N\kern.2em{\sc iii}]}

\begin{document}
\title{ISO LWS Spectroscopy of M82: A Unified Evolutionary Model}
 
\author{James W. Colbert\altaffilmark{1,12}, Matthew A. Malkan\altaffilmark{1}, Peter E. Clegg\altaffilmark{2}, Pierre Cox\altaffilmark{3}, Jacqueline Fischer\altaffilmark{4}, Steven D. Lord\altaffilmark{5}, Michael Luhman\altaffilmark{4,10}, Shobita Satyapal\altaffilmark{6,11}, Howard A. Smith\altaffilmark{7}, Luigi Spinoglio\altaffilmark{8}, Gordon Stacey\altaffilmark{9}, and Sarah J. Unger\altaffilmark{2}}
\altaffiltext{1}{Dept. of Physics \& Astronomy, University of California, Los
Angeles, CA 90095}
\altaffiltext{2}{Physics Department, Queen Mary \& Westfield College, University of London, Mile End Road, London E1 4NS, UK}
\altaffiltext{3}{Institut d' Astrophysique Spatiale, Bat. 120, Universite de Paris XI, F-91405 Orsay, France}
\altaffiltext{4}{Naval Research Laboratory, Remote Sensing Division, 4555 Overlook Avenue SW, Washington, DC 20375}
\altaffiltext{5}{Infrared Processing and Analysis Center, California Institute of Technology, Pasadena, CA 91125}
\altaffiltext{6}{Goddard Space Flight Center, Code 685, Greenbelt, MD 20771}
\altaffiltext{7}{Harvard-Smithsonian Center for Astrophysics, 60 Garden Street,
Cambridge, MA 02138}
\altaffiltext{8}{CNR-Istituto di Fisica dello Spazio Interplanetario, CP 27, 00044 Frascati, Italy}
\altaffiltext{9}{Dept. of Astronomy, 510 Space Science Bldg, Cornell University, Ithaca, NY 14853}
\altaffiltext{10}{NRC-NRL Research Associate}
\altaffiltext{11}{NRC-NASA Research Associate}
\altaffiltext{12}{email:colbert@astro.ucla.edu}

\setcounter{footnote}{0}
\begin{abstract}      

We present the first complete far-infrared spectrum (43 to 197 \m ) of M82, 
the brightest infrared galaxy in the sky, taken with the Long Wavelength Spectrometer of the Infrared Space Observatory (ISO). We detected seven fine structure emission lines, \oi \ 63 and 145 \m , \oiii \ 52 and 88 \m , \nii \ 122 \m , \niii \ 57 \m \ and \cii \ 158 \m , and fit their ratios to a combination starburst and photo-dissociation region (PDR) model. The best fit is obtained with HII regions with n = 250 cm$^{-3}$ and an ionization parameter of 10$^{-3.5}$ and PDRs with n = 10$^{3.3}$ cm$^{-3}$ and a far-ultraviolet flux of G$_o$ = 10$^{2.8}$. We applied both continuous and instantaneous starburst models, with our best fit being a 3-5 Myr old instantaneous burst model with a 100  M$_{\sun }$ cut-off. We also detected the ground state rotational line of OH in absorption at 119.4 \m . No excited level OH transitions are apparent, indicating that the OH is almost entirely in its ground state with a column density \ab 4x10$^{14}$ cm$^{-2}$. The spectral energy distribution over the LWS wavelength range is well fit with a 48 K dust temperature and an optical depth, $\tau _{Dust}$ $\propto$ $\lambda^{-1}$.      

\end{abstract}


\section{Introduction}

Infrared luminous galaxies emit an infrared luminosity comparable to or greater than their optical luminosity. The starburst galaxy is an infrared luminous galaxy converting its molecular ISM into stars at a rate that can not be sustained for a Hubble time. The radiation from these new stars is reprocessed into infrared radiation by the dust in their parental molecular clouds. Other possible energy sources for infrared luminous galaxies are active galactic nuclei or the energy of shocks resulting from galaxy interactions. Recent studies (e.g. Genzel et al. 1998) support the hypothesis that most of the brightest infrared luminous galaxies are predominantly powered by recently formed massive stars, however the simultaneous presence of AGNs and active star formation in some galaxies shows that both processes can occur in the same phase of the evolution of luminous infrared galaxies.

Due to its proximity (3.63 Mpc, Freedman et al. 1994) and moderate infrared luminosity, M82 is the brightest galaxy in the infrared. Because of its recent episode of star formation (e.g. Rieke et al. 1993) it is considered to be the prototypical starburst galaxy. Sometime around 10$^8$ years ago, M82 experienced a close encounter with M81 (Yun, Ho \& Lo 1993). This gravitational interaction likely produced the bar seen (Telesco et al. 1991), which provided the large streaming motions necessary to funnel large amounts of molecular gas to the galaxy's center. This large reservoir of molecular gas, \ab 2x10$^{8}$ M$_{\sun }$ (Wild et al. 1992), provides the fuel for the on-going starburst. 

Studies of the M82 nucleus in the mid-infrared (Telesco et al. 1991) saw strong off-center hot spots. Using the velocity profiles of the M82 far-infrared fine structure lines, Lord et al. (1996) modeled two hot spots as well. These hot spots may be star formation sites which result from cloud-cloud collisions that occur because of orbital crowding near the inner Lindblad resonances (Kenney et al. 1992). Alternatively, Satyapal et al. (1997) modeled the nucleus of M82 as a region of outward star propagation, with the hot spots being the most recent and brightest areas of star formation. 

Relatively insensitive to extinction, far-infrared spectroscopy can provide a unique probe of infrared-bright, dust obscured galaxies like M82. Line ratios may be used to constrain physical parameters and tend to be less vulnerable to calibration uncertainties as well. It is also in the far-infrared that some of the most important cooling lines of the ISM of galaxies, the \oi \ 63 \m \ and \cii \ 158 \m \ lines, appear. A close examination of M82 will provide a template for future comparisons to possible starburst galaxies, including those at high redshift. 

Section \S 2 describes our observations and the analysis used on the M82 far-infrared spectrum while section \S 3 presents our fine structure line results. Section \S 4 discusses line ratio model fitting, OH absorption and fitting of the entire spectrum before section \S 5 presents our final conclusions.  

\section{Observations and Analysis}
We present ISO LWS (Kessler et al. 1996, Clegg et al. 1996) grating mode (43-196.7\m , $\lambda $/$\Delta \lambda$ \ab 200) observations of M82. These observations were taken as part of the Long Wavelength Spectrometer Consortium's Extragalactic Guaranteed-Time Program. All observations were made in May 1996, on two separate orbits. The central position was R.A.(2000)=9$^{h}$55$^{m}$52$^{s}$.3 and Dec(2000)=69$^{o}$40'45.9''.  Our main observations consisted of 20 full grating scans, with a total integration over all scans of 8 seconds at each wavelength position spaced at one quarter of the resolution element. The spectral resolution was 0.29 \m \ for the 43-93 \m \ range and 0.6\m  \ for the 80-196\m \ range. Also, in the final averaged spectrum, the spectral resolution between 80 and 93 \m \ lies between these two values, resulting from our averaging of the overlap of two detectors that are detecting different orders of the grating. We also took a spectrum offset by $\Delta$\a =+0$^{s}$.8 and $\Delta$$\delta$=+17' from the center of the galaxy in order to measure the background. This offset observation had the same integration time, but was over a more limited wavelength range. All data used in this paper were processed through the LWS Pipeline Version 7.

The spectra were flux calibrated with respect to Uranus (Swinyard et al. 1998). The individual detector scans were calibrated to within $\pm$4-6$\%$ of each other, based on overlapping detectors, with the notable exception of detector LW4, covering wavelengths 150-170 \m , which was about 15\% \ lower. Several reliable detectors were chosen and the rest of the detectors were multiplicatively shifted so that overlapping points would match. The LWS beam is roughly 80'' FWHM, but does show some variation with detector, ranging from 65-85'' FWHM (Swinyard et al. 1998).

We performed the post-pipeline analysis with the ISO Spectral Analysis Package (ISAP). Even after pipeline processing the spectrum remains contaminated with bad data point `glitches' from cosmic ray hits, which we cleaned by plotting several scans on top of each other and looking for the characteristic fast rising and slowly falling `glitch' shape. All bad data points were then removed as well as any suspicious and widely deviating points following the glitch, where the detector showed memory effects. Roughly 20$\%$ of the data were discarded as unusable. 

Raw data also contain a fringing pattern that is believed to arise from interference between the beam coming off the field mirror and another beam reflected by the substrate holding that mirror. Fringing is only a minor effect for the shorter wavelengths ($<$70\m ): the fringing is weaker and the spectrum is inherently noisier. At wavelengths longer than 100\m , fringes approach \ab 5\% of the continuum, and endanger our ability to measure accurately the continuum level as well as identify possible weak emission and absorption lines. The fringe is sinusoidal in wavenumber with a well measured period of 3.54 cm$^{-1}$. We used the sinewave fitting algorithm for extended sources in ISAP to divide out the fringes. The defringing has little effect on the majority of line fluxes, altering the weak 145 \m \ and 122 \m \ lines by only 5\%. Only the OH absorption at 119 \m \ changes significantly, becoming 33\% \ weaker. Finally, we used a 3 $\sigma$ median clip on the data before averaging it all into a single spectrum with bins of 0.05 \m .
 
\section{Fine Structure Emission Lines}

Table 1 lists the seven fine structure lines detected along with 1 $\sigma $ uncertainties. The measured wavelength of each line is consistent with a 225 \kms \ redshift. The $\tau$'s listed are optical depths at each wavelength assuming a dust extinction model from Adams, Lada \& Shu (1988) and $\tau _{\lambda} \propto \lambda ^{-1}$ at wavelengths longer than \ab 50 \m . We calibrate this model using the visual extinction from the study of near infrared hydrogen recombination lines by Satyapal et al. (1995). Assuming a foreground screen they found an A$_V$ that ranged from 2-12 magnitudes, with its highest values near the nucleus where A$_V$\ab 10. We adopt this central value of 10 for our visual extinction, noting that at the wavelengths we are considering small errors in A$_V$ have little effect on the final line fluxes. The corrected flux is the measured flux multiplied by exp($\tau$), appropriate for an external dust screen. Figure 1 displays the entire spectrum. 

The fine structure lines we present have been measured in the past by
 the Kuiper Airborne Observatory (KAO) (Lord et al. 1996, Petuchowski et al. 1994, Stacey et al. 1991, Duffy et al. 1987, Lugten et al. 1986, Watson et al. 1984), but most were taken on separate nights with different calibrators. This complicates line ratios as beam sizes, spectral resolutions, and flux calibration schemes differ. Another advantage of this spectrum over KAO is the complete absence of telluric absorption features, which had greatest effect on the measurements of the \oi \ 63 \m \ and \nii \ 122 \m \ lines. We have improved the signal to noise on all lines, but the greatest increases were for the weak \nii \ 122 \m, \niii \ 57 \m \  and \oi \  145 \m \ lines. Table 2 gives the KAO lines and their measured continuum. Several of the KAO continuum measurements provided are published here for the first time.

The agreement between the KAO measurements and our own fluxes is satisfactory, with the most significant differences in the weak \nii \ 122 \m \ and \oi \ 145 \m \ lines. Our continuum measurements, however, go from being 30\% higher to 10\% lower than the KAO values. Some of this may be due to the smaller beams employed for KAO work (30 to 55''), but that can not explain all the discrepancies. 
      
The IRAS satellite measured a continuum flux of 1271 Jy at 60\m \ and 1351 Jy at 100\m \ (Rice et al. 1988), both lower measurements than those presented here. Since the IRAS values represent the total M82 flux in an area roughly 10x5 arcminutes, and since  the background measured by IRAS and by us in reference positions is less than 2 Jy, the reason for this discrepancy is unknown.

While the previous line work (Duffy et al. 1987, Lugten et al. 1986, Lord et al. 1996) achieved calibration accuracies to the 20-30$\%$ level, our study provides the first complete, medium resolution far-infrared spectrum, with substantially reduced uncertainty (\ab 5$\%$), no atmospheric problems, a standard aperature and a single calibration scheme. Furthermore, by using line ratios rather than absolute fluxes to determine starburst/PDR properties, most systematic uncertainties should cancel.

\section{Modeling and Other Results}

\subsection{Blackbody Fitting}

We fit our spectrum with a function of the form F$_{\lambda }$ $\propto$ B$_{\lambda }$(T)(1-e$^{-\tau _{Dust}}$), where we assume $\tau _{Dust}$ $\propto$ $\lambda ^{-1}$. This gave us a good fit for a temperature of 48 K for the emitting dust. If we then assume $\tau _{Dust}$ = 1 at \ab 2 \m \ (from A$_V$=10), we get a covering fraction of almost unity. Emissivity laws where $\tau \propto $ $\lambda^{-1.5}$ and $\lambda^{-2}$ were also considered. The $\lambda^{-1.5}$ law does not fit as well, while the $\lambda ^{-2}$ law showed large deviations. The 48 K temperature is consistent with the formula from Spinoglio et al. (1995), which determines the color temperature of a galaxy with a $\lambda ^{-1}$ emissivity law from its spectral index found using only the flux densities at 60 and 100 \m . From our data \a $_{60-100}$ for M82 is 0.04. When we substitute that into their regression formula, T$_{color}$=11.4 $\times$ (\a $_{60-100}$ +4.67)K, the result is 53K, only 5 degrees different than a fit to the entire far-infrared spectrum. 

Figure 2 shows the fit with extrapolation into the sub-mm. Not shown is Hughes, Gear \& Robson (1994)'s data point for 450 \m , $\lambda$F$_{\lambda }$ = 2.7\p 0.4x10$^{-17}$ W cm$^{-2}$, measuring the flux from the central 68''x68'' of M82. Inputting 450 \m \ into our fit gives a $\lambda$F$_{\lambda }$ = 5.5x10$^{-17}$ W cm$^{-2}$, a factor of 2 high, indicating, perhaps, a steepening in the $\lambda$ dependence for the dust emissivity law in the submillimeter. Klein, Wielebinski, \& Morsi (1988), fitting a few points from the infrared, submillimeter and out into the millimeter radio portion of the spectrum, did derive a steeper dust emissivity law, with $\tau _{Dust}$ $\propto$ $\lambda ^{-1.5}$.

The total flux for the infrared region observed is 8.15x10$^{-15}$ W cm$^{-2}$, which at a distance of 3.63 Mpc corresponds to 3.2x10$^{10}$L$_{\sun }$. Assuming it continues shortward of 43 \m \ as a black body, the total far infrared flux would be 9.7x10$^{-15}$ W cm$^{-2}$ or 3.8x10$^{10}$ L$_{\sun }$, in agreement with the estimate by Telesco $\&$ Harper (1980). The total flux in the fine structure lines is 6.0x10$^{-17}$ W cm$^{-2}$ or 0.6$\%$ of the infrared flux. The \cii /F(FIR) ratio in the LWS beam is 1.4x10$^{-3}$, similar to that seen toward galactic HII regions and other starburst and normal galaxies (Stacey et al. 1991).
 
\subsection{Emission Line Ratio Modeling}

A realistic starburst of a given age and duration is intermediate
between the simple cases of an instantaneous burst and continuous star formation. In order to limit the number of free parameters and to bracket the real situation, we looked at two different starburst types, instantaneous burst and continuous star formation. For each one our combined starburst HII region and PDR model has six free parameters: density in HII region, density in PDR, ionization parameter in HII region, far-ultraviolet flux in PDR, input starburst age and input starburst upper mass limit. 

Table 3 lists the measured line ratios of M82 together with the predicted combination starburst (HII region) and PDR model ratios. Following the methodology of Fischer et al. (1996), the HII region model was created using the program CLOUDY by Gary Ferland (Version 90-Ferland, 1996) and input spectral energy distributions (SEDs) of starbursts described in Leitherer \& Heckman (1995). To test different SED models, we tried two upper cut-off masses, 30 and 100 M$_{\sun }$ , and two burst types, instantaneous bursts and continuous star formation, for ages of 1-25 Myr. All the SED models had solar metallicity and initial mass function (IMF) slopes of 2.35. The input parameters are density and the ionization parameter, U, which is defined as the ratio of ionizing photons to hydrogen atoms at the inner face of the cloud:
\begin{equation}
U=\frac{Q(H)}{4\pi r^{2}n_{H}c}
\end{equation}
where Q(H) is the number of ionizing photons per second, n$_{H}$ is the total hydrogen density at the inner face of the cloud, and r is the distance from the ionizing source. Orion abundances and dust grains were assumed for the emitting HII regions. The \oiii \ 52 \m /\oiii \ 88 \m \ ratio is sensitive to density, but is almost completely independent of ionization parameter and model type. We found the best fit for a density of 250 cm$^{-3}$ for all models. The \niii 57\m /\nii \ 122 \m \ ratio acts in an orthogonal manner to our first diagnostic ratio. It is weakly dependent on density for low density HII regions, but is strongly dependent on ionization parameter and weakly dependent on model type. Here we fit log U varying from -3.1 for the 30 M$_{\sun }$ cut-off continuous star formation, to -3.5 for a 3-7 Myr old instantaneous burst with a 100 M$_{\sun }$ cut-off. These fits with different ionization parameters demonstrate Spinoglio \& Malkan (1992)'s point that a harder ionizing spectrum can be somewhat compensated for by a modest decrease in U. 

Change in input abundances made little difference in our derived starburst parameters. We ran CLOUDY models with Orion abundances, ISM abundances, solar abundances and twice solar metallicity starburst abundances, where the abundances are those defined in CLOUDY, except for the ISM abundances, which come from Sembach \& Savage (1996). No matter what abundances we input, we always derived basically the same densities and ionization parameters. The quality of the fit did vary somewhat, but this was mainly the fit of oxygen to nitrogen line ratios (\oiii \ 52 \m /\niii 57\m ,\oiii \ 88 \m /\nii \ 122 \m , etc.), which one would expect to change when altering the ratio of oxygen to nitrogen abundances. Orion abundances fit best, matching to within 10$\%$ of our measured ratios. ISM and solar abundances have only slightly different O/N abundance ratios and so they also produced satisfatory fits (within 30$\%$) of the measured O/N line ratios. Twice solar metallicity starburst abundances have high O/N abundance ratios and consequently produced high O/N line ratios, \ab 2.5$\times$ that observed. 

Of the four model input starburst SEDs considered, only two produced satisfactory ratio fits: the 3-7 Myr instantaneous burst with 100 M$_{\sun }$ \ cut-off and the 8-25 Myr continuous star formation with 30 M$_{\sun }$ \ cut-off. A fit was also achieved with the continuous star formation with 100 M$_{\sun}$ \ cut-off, but only after reducing the nitrogen abundance by a third. We can further constrain the models by looking at their total bolometric luminosity. Using the extinction-corrected Br$\gamma$ flux contained within an ISO LWS beam (from Satyapal et al. 1997) and assuming an electron density \ab 200 cm$^{-3}$ and an electron temperature \ab 5000K, we found an ionizing photon rate of 8x10$^{53}$ s$^{-1}$. This is similar to values previously found. For instance, McLeod at al. (1993) found a rate of 1.05x10$^{54}$ s$^{-1}$ using a smaller beam (30\arcsec ), but different assumptions about extinction. As one experiment, we scaled the Leitherer \& Heckman models to this ionizing photon rate and found that the continuous star formation produces \ab 10$^{11}$ L$_{\sun }$, more than twice the measured far-infrared luminosity (\ab 4x10$^{10}$ L$_{\sun }$). The older instantaneous bursts also produce too much luminosity, but the more recent bursts, 3-5 Myr old, only produce 3-5x10$^{10}$ L$_{\sun }$, more consistent with the observed total far-infrared luminosity. This makes the 3-5 Myr instantaneous burst our preferred input SED. Small errors in ionizing photon rate do not affect this conclusion. 

A similar check was made by comparing the Leitherer \& Heckman starburst masses expected to the dynamical mass derived from the mass model by Gotz et al. (1990), which gives us 1.6x10$^9$ M$_{\sun }$ within our ISO beam. McLeod et al. (1993) also used the Gotz et al. (1990) mass model, but for their smaller beam size. After they subtracted estimates for molecular mass and an older stellar population, they arrived at an estimate of 2.5x10$^{10}$ M$_{\sun }$ for the starburst population. Rieke et al. (1993) used this mass limit along with modeling to predict that a lower mass cut-off must exist. Our preferred instantaneous burst model predicts starburst masses of 0.5-1.3x10$^8$ M$_{\sun }$, depending on lower mass cut-off (1.0-0.1 M$_{\sun }$), which does not come close to exceeding the dynamical mass contained within the ISO beam. Even if we take the 2.5x10$^{10}$ M$_{\sun }$ used by Rieke et al. (1993) as our total allowed starburst mass, we have no trouble fitting the mass and see no requirement for a lower mass cut-off, although one might possibly exist.  

While the CLOUDY models fit the starburst lines well, they produce little \oi \ and \cii \ line emission. The amount of \cii \ 158 \m \ line flux produced by the CLOUDY models depends on the SED we input, but is \ab 24-31$\%$ of the total observed line flux for the instantaneous burst. Both \cii \ and \oi \ lines are strongly produced in PDRs, which one would expect to find in the interface between HII regions and molecular clouds. We use the PDR models of Kaufman et al. (1998). These models have two free parameters: density and far-ultraviolet flux, G$_o$, which is expressed in units of local Galactic far-ultraviolet flux or 1.6x10$^{-3}$ ergs s$^{-1}$ cm$^{-2}$. Once again the line ratios provide excellent diagnostics for determining these parameters. The ratio of \oi \ 63 \m /\oi \ 145 \m \ acts orthogonally to the \oi \ 63 \m /\cii \ 158 \m \ ratio over the densities and far ultraviolet fluxes considered (Wolfire, Tielens \& Hollenbach 1990). If we assume that the remaining 69-76$\%$ of the \cii \ not accounted for by the starburst HII region model is coming from the PDRs, then we have our PDR \oi \ 63 \m /\cii \ 158 \m \ ratio of \ab 1.6-1.8. Combining that with the observed \oi \ 63 \m /\oi \ 145 \m \ ratio (\ab 16) gives us a density of 10$^{3.3}$ cm$^{-3}$ and a G$_o$ around 10$^{2.8}$. These combined HII region and PDR models fit not only the line ratios, but, scaled to the distance and size of M82, they also fit the lines and continuum seen.  

We can check our derived G$_{o}$ and PDR density using another PDR diagnostic, the ratio of the PDR line fluxes to the infrared continuum flux. For our parameters the Kaufman et al. (1998) models give Flux(CII158 \m+CII63 \m+OI145 \m)/FIR=3.9x10$^{-3}$, roughly equal to the 2.9x10$^{-3}$ we get after subtracting our HII region line contribution. It should be noted that the Kaufman et al. (1998) PDR models use Galactic ISM abundances, not the Orion abundances input into our CLOUDY HII region models. However, for our PDR density and G$_{o}$, the difference in using two different abundances is negligible.

Since the regions we are modeling are in physical contact with one another, we should be able to make some continuity checks across the model boundaries. Following the treatment described by Satyapal et al. (1998), we start by checking that there is pressure equilibrium between the HII region and the PDR. The edge of the HII region has a temperature of 4000 K, which gives a P/k \ab \ 10$^6$ cm$^{-3}$ K. Inputing our PDR surface temperature of \ab 250K, derived from Kaufman et al. (1998), and density gives a P/k \ab \ 5x10$^5$ cm$^{-3}$ K or roughly the same as our HII region. The second boundary condition is that the far ultraviolet flux leaving the HII region should equal that entering our PDR model. We measure the far ultraviolet flux leaving the HII region by taking the CLOUDY output continuum and integrating the flux from 6 to 13 eV. This depends strongly on input SED, giving us G$_o$=10$^{2.2}$-10$^{2.4}$ for the instantaneous burst models. Considering the uncertainties, this is consistent with our model G$_o$=10$^{2.8}$.

Lester et al. (1987) noted that the \oiii \ 52 \m /\niii \ 57 \m \ ratio will equal the actual O$^{++}$/N$^{++}$ ratio within $\pm$50$\%$, thanks to the similar critical densities for thermalization of their respective emitting levels and ionization potentials that are within 25$\%$ of each other.  With knowledge of the electron density, which one can get accurately from the \oiii \ 52 \m /\oiii \ 88 \m \ ratio, a very precise O$^{++}$/N$^{++}$ ratio can be found that holds true for a broad range of electron temperatures. Following the calculations of Lester et al. (1987) and taking our density of 250 cm$^{-3}$ we find a O$^{++}$/N$^{++}$ ratio of 3.8. This is slightly higher than that reported by Duffy et al. (1987) for M82, O$^{++}$/N$^{++}$=3.1. Both ratios are higher than the average value of HII regions towards the Galactic Center, where $<$ O$^{++}$/N$^{++} >$ \ab \ 0.9-1.4 (Dinerstein et al. 1984). One possible conclusion that could be drawn is that the O/N ratio is truly higher in M82. However, it is more likely that this is an effect of the local ionizing conditions. In low-ionization conditions, i.e. cool stars, more nitrogen is doubly ionized than oxygen, because the ionization potential of N+ is smaller than that of O+. The difference in O$^{++}$/N$^{++}$ ratio in the Galaxy would result from less massive (cooler) stars creating the HII regions. This is further supported by our model fit, where we fit the O$^{++}$ and N$^{++}$ lines for M82 with a high-ionization starburst SED and Orion abundances without any oxygen over-abundance. In fact, as this section mentioned earlier, the more oxygen-rich abundance models produced significantly worse fits to our line ratios.

Examining our best fit we come within 10$\%$ of all the ratios and closer on most. The derived parameters: HII region density, PDR density, ionization parameter, far-ultraviolet flux, burst upper mass cut-off and burst age, are well determined, with even small changes in any of these inputs creating ratios outside the uncertainties of the ratios measured. The majority of error still present in these fits comes from the models themselves and the assumptions we made, not from the inaccuracy in our line ratio measurements.

We can compare this work with previous models. Spinoglio \& Malkan (1992) also fit far-infrared lines with a combination HII region and PDR model. While achieving similar HII region densities, their M82 model had a much larger ionization parameter (10$^{-2.5}$). This is entirely due to the hardness of the input SED. The Leitherer \& Heckman (1995) SEDs are significantly harder than the SED Spinoglio \& Malkan (1992) used. The Spinoglio \& Malkan (1992) PDR parameters were also different (G$_o$=10$^{4}$, n=10$^{4-5}$ cm$^{-3}$), but this can be mainly attributed to the newer PDR models of Kaufman et al. (1998). The Lord et al. (1996) PDR parameters for M82 (G$_o$=10$^{3}$, n=10$^{4}$ cm$^{-3}$) come closer to our derived values, but also differ because of the newer models used. 

Satyapal et al. (1997) and Reike et al. (1993) previously created models of the M82 starburst, both finding typical starburst ages of around 10$^7$ years. Reike et al. (1993) compared many observational parameters (M$_{K}$, mass, luminosity, CO index, UV photons), fitting M82 with two separate starbursts, the first being 13-30 Myr ago. Satyapal et al. (1997) looked at individual starburst clusters, measuring their CO indices and Br$\gamma$ equivalent widths. Modeling those features for each individual cluster, they found a range in age of 6x10$^6$ years, implying an outward propagating starburst. For the sake of simplicity, we did not attempt to combine multiple models with different starburst ages, but we note many of the indivual starburst cluster ages found by Satyapal et al. (1997) were around 5 Myr. 

One unknown that could throw off the models is the degree of \oi \ 63 \m \ self-absorption in M82. Stacey et al. (1983) first observed the 145 \m \ \oi \ line from the PDR associated with the Orion A HII region, and discovered that the \oi \ 63, 145 and \cii \ 158 \m \ line intensity ratios indicated optically thick ($\tau $\ab 2) emission in the 63 \m \ line. Tielens and Hollenbach (1985) models also indicated that this should be so. Later, through more sophisticated modeling involving the [$^{13}$CII] line, and more complete mapping in the \cii \ and \oi \ lines,  Stacey et al. (1993) estimated that the \oi \ 63 \m \ has an optical depth \ab \ 3 in the Orion PDR. The \oi \ 63 \m \ line is seen in absorption in Arp 220 (Fischer et al. 1998) and in Galactic star formation regions (Poglitsch et al. 1996, Baluteau et al. 1997). In addition, Hermann et al. (1997) explained an unusually low \oi \ 63 \m /\oi \ 145 \m \ ratio in the Dark Lane region of the Orion Molecular Cloud as a likely site of \oi \ 63 \m \ self-absorption. 

One would expect \oi \ 63 \m \ absorption to occur where cold foreground oxygen lies between us and an emitting source. However, the gas density must exceed the critical density for collisional de-excitation ($n_{crit} \sim 5\times 10^5$ cm$^{-3}$) in order for the absorbed photons to be removed from our beam. At
lower densities the photons are mainly scattered with little reduction
in the observed line luminosity.  In our Galaxy, such dense clouds
rarely cover more that a small fraction of the \oi \  emitting
region. Since the ISM in most starburst and normal galaxies seem to be constructed of a superposition of many such clouds which usually do not cover each other simultaneously in both velocity and area, the small regions of \oi \ self absorption should not be important. A notable exception is Arp 220 (Fischer et al. 1998), where it is likely the FIR source is largely enveloped by cold, dense clouds with abundant neutral oxygen. There is no reason to think M82 such an extreme case and in fact, our analysis of high signal-to-noise Fabry-Perot scans of the \oi \ 63 \m \ line does not indicate any self-absorption. Any significant \oi \ self-absorption would alter the PDR modeling by increasing the PDR density.
   
There is one more important forbidden line in the far infrared, the [NII]205 \m \ line, which lies just beyond the LWS limit of 196.7 microns. Petuchowski et al. (1994) measured it using the KAO and our model does predict its relative flux. We include both on our line ratio table and note they match to within the uncertainty of the KAO flux measurements. The [NII] line ratio is the best density tracer for low density (n $<$ 500cm$^{-3}$) HII regions, so the accurate reproduction of this line ratio lends credence to our estimate: n$_e$ \ab \ 250 cm$^{-3}$. 

\subsection{Cross scans of the CII and OI lines}

In addition to the full scan made at the center of M82, we collected cross-scans of the \cii \ 158 \m \ and \oi \ 63 \m \ lines using the LWS02 AOT. The maps consist of two cuts across M82, one across the major axis with a P.A.=55$^o$ and the other across the minor axis, perpendicular to the first. Seven observations were made in each cut, starting 150 arcseconds from M82's center and spaced 50 arcseconds apart. Each line represents an integration of 3 seconds per spectral element. The two plots in Figure 3 show the relative strengths of the emission lines along the two axes. Only upper limits were found for the \oi \ 63 \m \ line at the points 150 arcseconds from M82's center and therefore are not included on the map. The major axis cut runs from southwest to northeast, while the minor axis cut runs from northwest to southeast. 
            
Both lines weaken away from the center, but while the \cii \ 158 \m \ line fall-off is symmetrical about the minor axis, it is not symmetrical about the major axis and the \oi \ 63 \m \ line is not symmetrical around either axis. This asymmetry is not that surprising. Previous maps, such as those in H\a \ (Waller, Gurwell \& Tamura 1992), the near infrared (Satyapal et al. 1995) and the submillimeter (Hughes, Gear \& Robson 1994), show an excess in flux to the west. Stacey et al. (1991) did similar cuts in the \cii \ 158 \m \ line which also showed this slight asymmetry in the major axis. Lord et al. (1996) did detailed modeling of \oi \ 63 \m \ line profile, attributing the majority of the line flux to two off-center hot spots, with the western spot the brighter of the two. We examined the far infrared continuum in our cross scans and also found more flux to the west. 

We measured the spectral index \a \ between 60 and 100 \m , where F$_{\nu }$ $\propto$ $\nu ^{\alpha }$, at every spot in our scans. This \a $_{60-100}$ is largest 50 arcseconds to the west, where it is \ab 0.4, not at M82's center, where \a $_{60-100}$ is \ab 0.1. This higher \a $_{60-100}$ implies a higher average temperature, further indication of a strong hot spot to the west. In addition, we also looked at the spectral index \a \ between 100 and 175 \m, where we were interested in testing whether our single temperature black body with emissivity $\propto$ $\lambda ^{-1}$ would fit further outward from the galaxy's center. The \a $_{100-175}$ at the hot spot to the west is \ab 1.6, which is consistent with our black body fit. However, the \a $_{100-175}$ measured to the east and farther outwards in the galaxy is smaller than expected, possibly indicating a growing cold dust component.  

Most interesting is the variation in the \oi \ 63 \m/\cii \ 158 \m \ line ratio, visible in Figure 4. Assuming the two lines come mostly from PDRs, tracing the \oi \ 63 \m/\cii \ 158 \m \ line ratios will tell us something about conditions in M82 on a large scale. The ratio shows a maximum roughly 50 arcseconds off center to the northwest and southwest of the galaxy's center and shows a steady decrease outwards. According to the models of Kaufman et al. (1998), this ratio decreases with PDR density and far ultraviolet field strength. The general trend of decreasing \oi \ 63 \m/\cii \ 158 \m \ ratio with distance from M82's center is then understandable. One would expect both density and far ultraviolet field to decrease with distance from the galaxy's center. The off-center maximum in the \oi \ 63 \m/\cii \ 158 \m \ ratio must indicate where conditions of high density, high G$_o$ or both exist. 

The \cii \ 158 \m \ line flux also appears more extended than the \oi \ 63 \m \ line flux, possibly indicating its association with the filamentary features observed in optical recombination lines as previously suggested by Stacey et al. (1991). Taking the formula for \cii \ intensity originating from ionized regions from Madden et al. (1993) we get
\begin{equation}
I_{C+}(ionized) = \frac{h\nu }{4\pi }\frac{A}{n_{crit}}\frac{g_u/g_l}{1 + [(g_u/g_l) + 1](n_e/n_{crit})}X_{C+}EM
\end{equation}
where A is the spontaneous emission coefficient for the $^2$P$_{3/2}$-$^2$P$_{1/2}$ transition (2.36x10$^{-6}$ s$^{-1}$), $g_u/g_l$ is the ratio of statistical weights in the upper and lower levels (2), $n_{crit}$ is the critical density (\ab 35 cm$^{-3}$), $X_{C+}$ is the abundance of C$^+$ relative to hydrogen (\ab 3.3x10$^{-4}$) and EM is the emission measure. Assuming electron density is low compared to the critical density, we only need a measurement of the emission measure to derive the expected flux from the extended, diffuse ionized gas of M82. We derive the EM from the 6 cm maps of Seaquist \& Odegard (1991), assuming the majority of the emission comes from thermal free-free emission. At roughly 50 arcseconds off the minor axis we find an EM \ab 1300 cm$^{-6}$ pc, giving us an expected intensity for C$^+$ of 1.7x10$^{-4}$ ergs sec$^{-1}$ cm$^{-2}$ sr$^{-1}$ or an expected flux in our ISO LWS beam of 2.0x10$^{-18}$ W cm$^{-2}$. This number is consistent with that observed, indicating that the majority of the extended C$^+$ emission is coming from the diffuse ionized medium. 
            
\subsection{OH absorption}

The only certain absorption feature observed in the spectrum of M82 is due to the two lambda-doubling rotational transitions of OH at \ab 119.4 \m , which are unresolved at the LWS grating-mode resolution. These are the transitions from the OH ground state of $^{2}\pi _{3/2}$ J=3/2 to the next highest energy state, $^{2}\pi _{3/2}$ J=5/2, which have been seen before in emission in the KL nebula (Storey, Watson \& Townes 1981) and in absorption toward the galactic center (Genzel et al. 1985). No other OH lines are observed in M82. There is a dip at the location of the 53 \m \ transitions, but it is the same scale as the noise and must be treated as an upper limit. Table 4 lists the 1 $\sigma $ upper limits for the next most likely OH lines. 

The equivalent width of an optically thin line is
\begin{equation}
W=A_{ij}\frac{g_i}{g_j}\frac{\lambda _{o}^4}{8\pi c}N
\end{equation}
where A$_{ij}$ is the Einstein coefficient for spontaneous emission between two levels, i and j, g is their statistical weights, $\lambda _{o}$ is the transition's line center and N is the column density of material (Aller 1960). Taking the A-coefficients from Destombes et al. (1977), one can calculate column densities and column density upper limits for the lines. Three of the transitions, 53 \m , 79 \m \ and 119 \m, are all from the ground state, but we only detect the 119 \m \ line because the A-coefficients for 53 \m \ and 79 \m \ give equivalent widths 20 to 80 times less than that of the 119 \m \ transition, assuming the transitions are optically thin. This is consistent with the upper limits observed. The lack of the 84 \m \ and 163 \m \ lines, despite having higher cross sections, demonstrates that the absorbing OH gas is almost entirely in its ground state. The column density of OH in its ground rotational state is \ab \ 4.2x10$^{14}$ cm$^{-2}$. Further analysis will be presented in a later paper.

Besides the 119 \m \ OH absorption lines and the seven emission lines, we identify no other lines or other features in the spectrum. We should, however, point out some visible features in the spectrum that may or may not be real. The large bumps at the shortest wavelengths, 43-50 \m , are present in other LWS spectra (Arp 220, Fischer et al. 1998). These features appear to be an artifact of that detector, SW1, but they are being investigated further. Two other apparent features at 56 and 112 \m, result from unexpected HD lines in the Uranus calibration spectrum. The 56 \m \ line is particularly troublesome, as it confuses the situation on the short wavelength side of the \niii 57\m \ line, where there is a hint of an absorption, possibly a p-H$_2$O line. Harvey et al (1998) have examined the ISO/LWS M82 spectrum for evidence of atomic hydrogen recombination lines in this range.  None were detected. The upper limits, constrained by their very low line-to-continuum ratio,  lie within the range of expected values set by the known centimenter and decimeter lines and a spontaneous emission
process. Other potential features not already mentioned are absorptions at 65.1 \m , 66.4 \m \ and  149.2 \m . Further analysis and possible identification will come in a later paper. Any undetected lines present in the M82 spectrum would have to have fluxes less than \ab 5x10$^{-19}$ W cm$^{-2}$ at 50 \m \ to \ab 2x10$^{-19}$ W cm$^{-2}$ at 180 \m. 

\section{Conclusions}

We have obtained an LWS full grating scan of the prototypical starburst galaxy, M82. This spectrum reveals seven bright fine structure lines, previously detected from the KAO, but here detected at the same time, with the same instrument, with roughly the same beam and without any obstructing atmosphere. Therefore, the relative calibration of these lines is superior to that used for prior analysis. Also reported is the discovery of an absorption line of OH from its lowest state. The upper limits for other possible OH lines are enough to determine that most of the absorbing OH is in its ground state. The whole spectrum is well fit by a single temperature dust component with an optical depth $\tau _{Dust}$ $\propto$ $\lambda ^{-1}$, giving a total infrared flux of 3.8x10$^{10}$ L$_{\sun }$. 

We fit the infrared line ratios of M82 with a six parameter combination HII region and PDR model. The parameters for the best fitting model are: HII region density = 250 cm$^{-3}$, U = 10$^{-3.5}$, PDR density = 10$^{3.3}$ cm$^{-3}$, G$_o$=10$^{2.8}$, upper mass cut-off = 100M\sun \ and age = 3-5 Myr. The \cii \ 158 \m \ line comes \ab 25$\%$ from the HII regions and \ab 75$\%$ from PDRs. In our model, the starburst mass is \ab 0.5-1.3x10$^8$ M$_{\sun }$, depending on the lower mass cut-off chosen. This is a significant percentage of the molecular mass (\ab 2x10$^{8}$ M$_{\sun }$, Wild et al. 1992), indicating the M82 starburst will not be able to continue producing stars, through another instantaneous burst or otherwise, for much longer before running out of mass. The LWS cross-scan data support the model of strong off-center hot spots seen in the mid-infrared (Telesco et al. 1991) and modeled from velocity profiles by Lord et al. (1996). Our fit for a 3-5 Myr instantaneous starburst is a simplification of the true situation, which is several regions with different starburst ages. The single burst we modeled should be dominated by the brightest and most recent bursts, which are presumably the hot spots themselves.   

This work was made possible by the hard work and dedication of the ISO
team and the LWS consortium.  We gratefully acknowledge the use of the
Kaufman, Wolfire, Hollenbach, and Luhman (1998) models. In particular we would like to thank Matt Greenhouse, Mark Wolfire, and Michael Kaufman for their comments and help. This work was supported by the NASA ISO program, including NASA grants NAG 5-3309, NAGW-1711 and NAGW-4038. Jacqueline Fischer and Michael Luhman also acknowledge support by the Office of Naval Research.

\pagebreak

\begin{figure}
\centering \leavevmode
\epsfxsize=1.0\columnwidth
\epsfbox{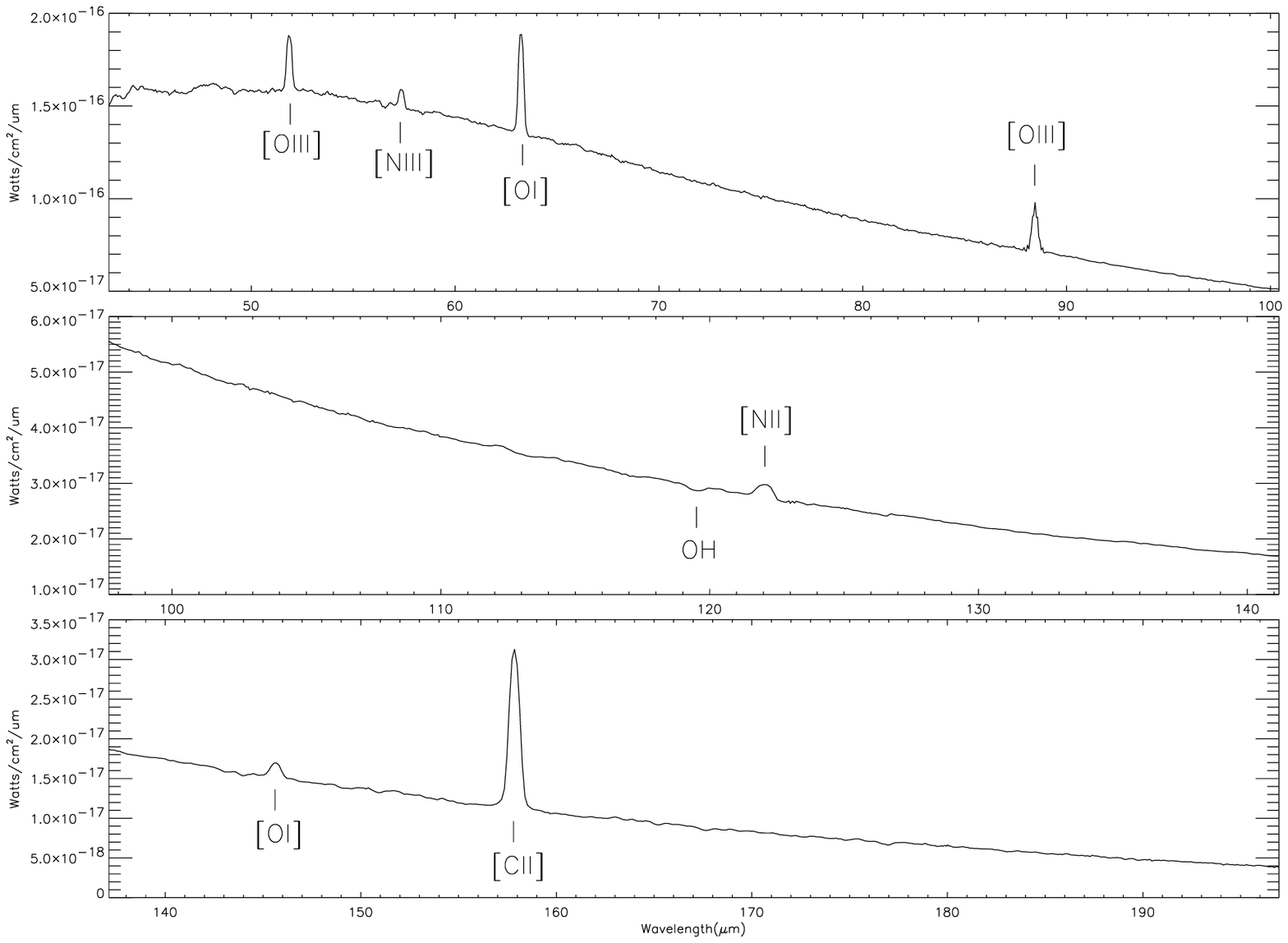}
\figcaption{LWS spectrum for M82. Seven fine structure lines are visible, as well as an OH absorption at 119 \m . The strong features seen around 43-50\m \ are unidentified and possibly artifacts of the first detector.}
\end{figure}

\begin{figure}
\centering \leavevmode
\epsfxsize=1.0\columnwidth
\epsfbox{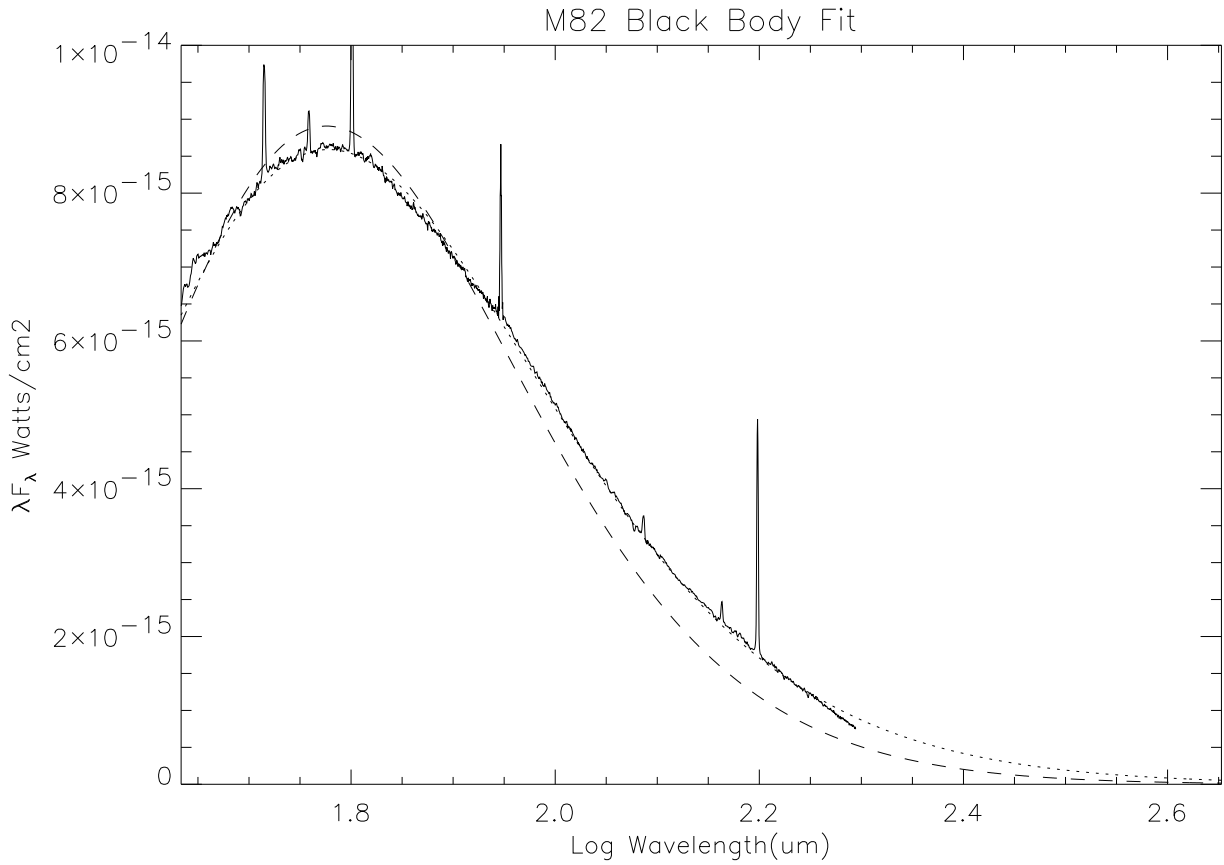}
\figcaption{Blackbody Fit. Solid line is actual data corrected for predicted extinction. Dashed line is best fit black body spectrum with flux density $\propto$ \ B$_{\lambda }$(T)(1-e$^{-\tau _{Dust}}$) law, where $\tau \propto \lambda ^{-1}$. For comparison, the dotted line is our best fit with $\tau \propto \lambda ^{-2}$.}
\end{figure}

\begin{figure}
\centering \leavevmode
\epsfxsize=1.0\columnwidth
\epsfbox{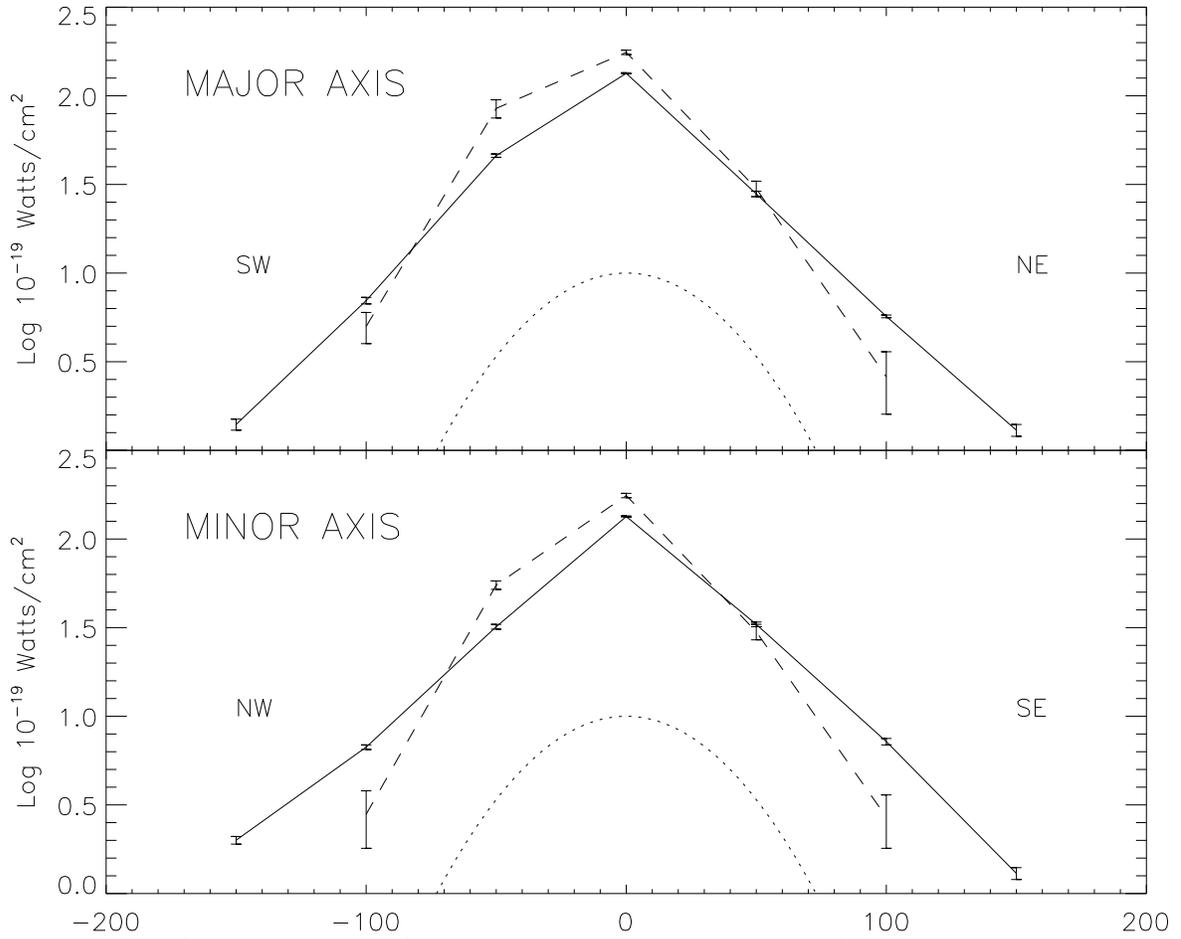}
\figcaption{Plots of the strengths of two lines across the major (top) and minor (bottom) axis of M82. The solid line is the \cii \ 158 \m \ line while the dashed line is the \oi \ 63 \m \ line. The LWS beam is represented by the dotted line.}
\end{figure}

\begin{figure}
\centering \leavevmode
\epsfxsize=1.0\columnwidth
\epsfbox{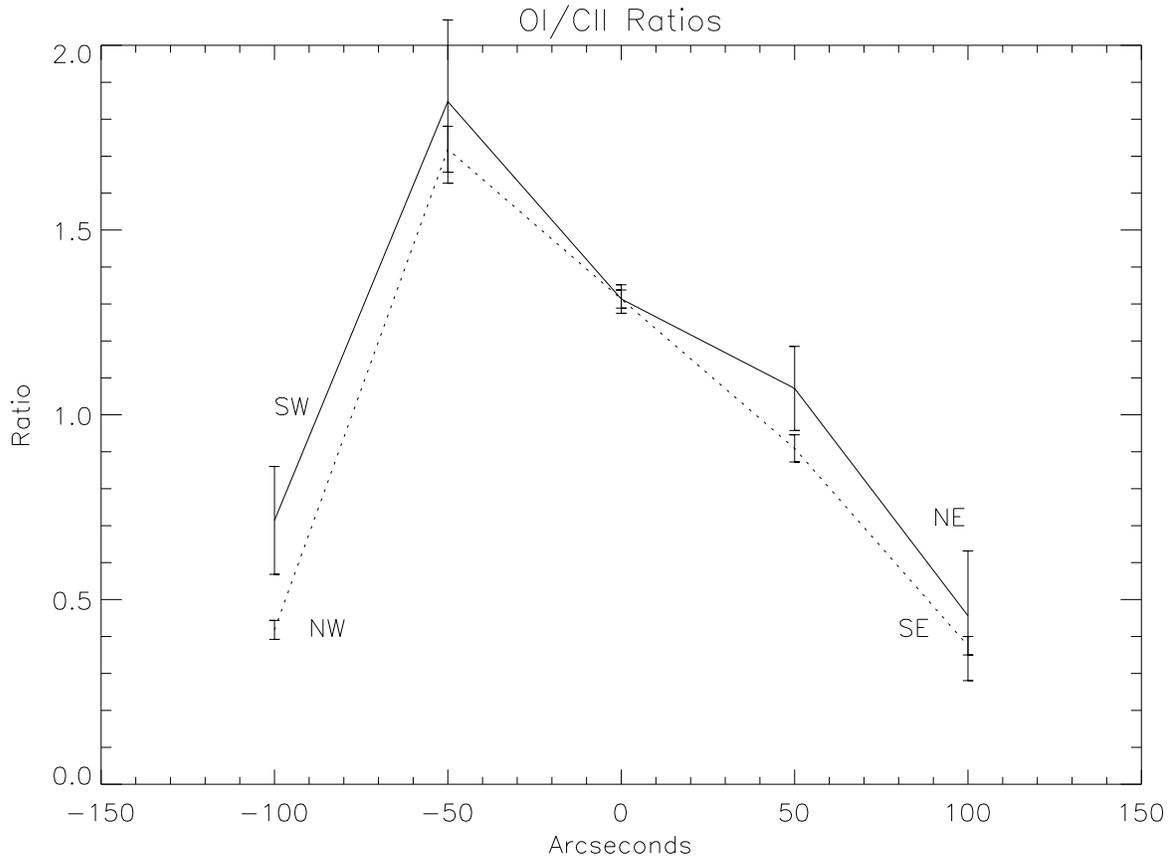}
\figcaption{Plot of \oi /\cii \ line ratio across M82. The solid line is the ratio across the major axis while the dashed line is the ratio along the minor axis.}
\end{figure}

\begin{table}
\begin{tabular}{|lccccc|}
\multicolumn{6}{c}{\bf Table 1} \\
\multicolumn{6}{c}{\bf Emission Line Fluxes} \\
\multicolumn{6}{c}{\hspace*{3cm}} \\ \tableline
Line & Rest $\lambda$(\m ) & Flux(W cm$^{-2}$) & $\tau$ & Corr. Flux(W cm$^{-2}$)\tablenotemark{a} & Cont.(Jy)\tablenotemark{b} \\ \tableline
\ \oiii & 51.81  & 10.3$\pm 0.5$ x 10$^{-18}$ & 0.10 & 11.3$\pm 0.6$ x 10$^{-18}$& 1490  \\
\ \niii & 57.32 & 3.4$\pm 0.5$ x 10$^{-18}$ & 0.09 & 3.7$\pm 0.5$ x 10$^{-18}$ & 1695  \\
\ \oi  & 63.18 & 17.6$\pm 0.5$ x 10$^{-18}$ & 0.08 & 19.1$\pm 0.5$ x 10$^{-18}$ & 1860 \\
\ \oiii & 88.36 & 8.6$\pm 0.4$ x 10$^{-18}$ & 0.06 & 9.1$\pm 0.4$ x 10$^{-18}$ & 1970 \\
\ \nii & 121.90 & 1.7$\pm 0.3$ x 10$^{-18}$ & 0.04 & 1.8$\pm 0.3$ x 10$^{-18}$ & 1460 \\
\ \oi & 145.53 & 1.2$\pm 0.1$ x 10$^{-18}$ & 0.04 & 1.2$\pm 0.1$ x 10$^{-18}$ & 1155 \\
\ \cii & 157.74 & 13.4$\pm 0.1$ x 10$^{-18}$ & 0.03 & 13.8$\pm 0.1$ x 10$^{-18}$ & 1005  \\  \tableline
\end{tabular}
\tablenotetext{a}{Corrected flux is the measured flux corrected for extinction.}
\tablenotetext{b}{Contiuum flux density has not been corrected for extinction.}
\end{table}

\begin{table}   
\begin{tabular}{|lcccc|}
\multicolumn{5}{c}{\bf Table 2} \\
\multicolumn{5}{c}{\bf KAO Continua and Fluxes} \\
\multicolumn{5}{c}{\hspace*{3cm}} \\ \tableline
Line & KAO Line(W cm$^{-2}$)\tablenotemark{a} & KAO Continuum(Jy) & HPBW($\arcsec$) & Reference \\ \tableline
\ \oiii 52 \m & 9.5\p 0.7 x 10$^{-18}$ & 1207 & 48 & 1 \\
\ \niii 57 \m & 3.9\p 0.4 x 10$^{-18}$ & 1269 & 48 & 1 \\
\ \oi \ 63 \m & 14.2\p 3.4 x 10$^{-18}$ & 1254 & 44 & 2 \\ 
\ \oiii \ 88 \m & 8.6\p 0.5 x 10$^{-18}$ & 1689 & 48 & 1 \\
\ \nii \ 122 \m & 2.9$^{+0.9}_{-0.6}$ x 10$^{-18}$ & 1190 & 45 & 3 \\
\ \oi \ 145 \m & 0.84\p 0.24 x 10$^{-18}$ & 1130 & 55 & 4 \\
\ \cii \ 158 \m & 14\p 4.2 x 10$^{-18}$ & 1150 & 55 & 4 \\  \tableline
\end{tabular}
\tablenotetext{a}{All line fluxes have been corrected for same extinction used in Table 1.}
\tablerefs{(1)Duffy et al. 1987; (2)Lord et al. 1996; (3)Petuchowski et al. 1994; (4)Lugten et al. 1986.}
\end{table}

\begin{table}   
\begin{tabular}{|lccc|}
\multicolumn{4}{c}{\bf Table 3} \\
\multicolumn{4}{c}{\bf Line Ratios} \\
\multicolumn{4}{c}{\hspace*{3cm}} \\ \tableline
Line Ratio & Measured Ratio & Inst. Model\tablenotemark{a} & Cont. Model\tablenotemark{b} \\ \tableline
\ \oiii \ 52 \m /\niii 57\m  & 3.0\p 0.4 & 3.0 & 3.1 \\
\ \oiii \ 52 \m /\oiii \ 88 \m  & 1.24\p 0.08 & 1.26 & 1.32 \\
\ \oiii \ 52 \m /\oi \ 63 \m  & 0.59\p 0.04 & 0.59 & 0.59 \\
\ \oiii \ 52 \m /\nii \ 122 \m  & 6.3\p 0.5 & 6.3 & 6.6 \\  
\ \oiii \ 52 \m /\oi \ 145 \m  & 9.4\p 0.9  & 9.2 & 10.5 \\
\ \oiii \ 52 \m /\cii \ 158 \m  & 0.82\p 0.04 & 0.82\tablenotemark{c} & 0.82\tablenotemark{c} \\
\ \niii 57\m /\nii \ 122 \m  & 2.1\p 0.3 & 2.1 & 2.1 \\
\ \oi \ 63 \m /\oi \ 145 \m  & 15.9\p 1.4 & 15.5 & 17.9 \\
\ \oi \ 63 \m /\cii \ 158 \m  & 1.38\p 0.03 & 1.39 & 1.41 \\ 
\ \nii \ 122 \m /[NII]205 \m & 4.1 \p 1.4\tablenotemark{d} & 4.4 & 4.9 \\ \tableline
\end{tabular}
\tablenotetext{a}{Instantaneous Burst, 3-5 Myr old with 100 M$_{\sun }$ cut-off + PDR contribution. The preferred model.}
\tablenotetext{b}{Continuous Star Formation, 8-25 Myr old with 30 M$_{\sun }$ cut-off + PDR contribution. This model was rejected because it produced too much luminosity.}
\tablenotetext{c}{This ratio is forced to exactly match by the fitting method.}
\tablenotetext{d}{ISO did not observe the [NII]205 \m \ line; we use the KAO observation from Petuchowski et al. (1994).} 
\end{table}

\begin{table}    
\begin{tabular}{|lccc|}
\multicolumn{4}{c}{\bf Table 4} \\
\multicolumn{4}{c}{\bf OH lines} \\
\multicolumn{4}{c}{\hspace*{3cm}} \\ \tableline
Line & Transition & Equivalent Width(\m ) & A$_{ij}$(sec$^{-1}$) \\ \tableline
53 \m & $^{2}\pi _{3/2} J=3/2 \rightarrow \ ^{2}\pi _{1/2}$ J=3/2 & $<$ 0.003 & 0.04 \\
79 \m & $^{2}\pi _{3/2} J=3/2 \rightarrow \ ^{2}\pi _{1/2}$ J=1/2 & $<$ 0.004  & 0.033 \\
84 \m & $^{2}\pi _{3/2} J=5/2 \rightarrow \ ^{2}\pi _{3/2}$ J=7/2 & $<$ 0.004 &  0.5 \\  
119 \m & $^{2}\pi _{3/2} J=3/2 \rightarrow \ ^{2}\pi _{3/2}$ J=5/2 & 0.022\p 0.004 & 0.13 \\
163 \m & $^{2}\pi _{1/2} J=1/2 \rightarrow \ ^{2}\pi _{3/2}$ J=3/2 & $<$ 0.005  & 0.054 \\ \tableline
\end{tabular}
\end{table}

\end{document}